\documentclass[11pt,a4paper]{article}
\pdfoutput=1
\usepackage{a4wide}
\usepackage{breakcites}

\usepackage{booktabs}
\usepackage{graphicx,color,epstopdf}
\usepackage{hyperref}
\usepackage{amsmath,amssymb}
\usepackage{hyperref}
\usepackage{array}

\usepackage[small,bf]{caption}
\usepackage{color}
\usepackage[dvipsnames]{xcolor}
\usepackage{cite}
\usepackage{bbm}
\usepackage{slashed}
\usepackage{tcolorbox}
\usepackage{mathtools}
\usepackage{braket}

\usepackage[utf8]{inputenc}
\usepackage{amsmath}
\usepackage{amssymb}
\usepackage{graphicx}
\usepackage{multirow}

\begin{document}


\begin{center}
{ \bf\LARGE
A Modern Look at the Oscillation Physics Case for a Neutrino Factory
}
\\[8mm]
Peter B.~Denton$^{\, a,}$ \footnote{E-mail: \texttt{pdenton@bnl.gov}} 
and Julia Gehrlein$^{\, b,}$ \footnote{E-mail: \texttt{julia.gehrlein@colostate.edu}} 
\\[1mm]
\vspace*{0.50cm}
{\textit{\small $^{a}$High Energy Theory Group, Physics Department, Brookhaven National Laboratory, Upton, NY 11973, USA}\par}
{\textit{\small $^{b}$Physics Department, Colorado State University, Fort Collins, CO 80523, USA}\par}
\vspace*{1.2cm}
\end{center}


\begin{abstract}
The next generation of neutrino oscillation experiments, JUNO, DUNE, and HK, are under construction now and will collect data over the next decade and beyond.
As there are no approved plans to follow up this program with more advanced neutrino oscillation experiments, we consider here one option that had gained considerable interest more than a decade ago: a neutrino factory.
Such an experiment uses stored muons in a racetrack configuration with extremely well characterized decays reducing systematic uncertainties and providing for more oscillation channels.
Such a machine could also be one step towards a high energy muon collider program.
We consider a long-baseline configuration to SURF using the DUNE far detectors or modifications thereof, and compare the expected sensitivities of the three-flavor oscillation parameters to the anticipated results from DUNE and HK.
We show optimal beam configurations, the impact of charge identification, the role of statistics and systematics, and the expected precision to the relevant standard oscillation parameters in different DUNE vs.~neutrino factory configurations.
\end{abstract}

\section{Introduction}
The next generation of neutrino oscillation experiments, including JUNO \cite{JUNO:2022mxj}, DUNE \cite{DUNE:2020ypp}, and HK \cite{Hyper-KamiokandeProto-:2015xww}, are expected to make significant headway in addressing three of the remaining known unknowns in particle physics.
Notably, they will have excellent $\gg5\sigma$ sensitivity to the mass ordering (MO) (the sign of $\Delta m^2_{31}$), they will have good $\theta_{23}$ octant determination capabilities provided that it isn't too close to maximal, and both DUNE and HK will each individually provide $>5\sigma$ sensitivity to CP violation for significant regions of parameter space.
In addition, they will provide some resolution on measuring $\delta$ (we use the standard parameterization of the PMNS \cite{Pontecorvo:1957cp,Maki:1962mu} matrix \cite{Denton:2020igp}).
Atmospheric neutrino experiments such as HK, the IceCube-upgrade \cite{Ishihara:2019aao}, and KM3NeT-ORCA \cite{KM3Net:2016zxf} will also provide complementary information.
For a recent overview of this picture, see \cite{Denton:2022een}.

While the outlook for determining the neutrino oscillation parameters is hopefully rosy, it is important to carefully examine what, if anything, should come after these experiments in the neutrino oscillation program.
Such an experiment would provide high precision measurements of all the oscillation parameters\footnote{The solar parameters will not be well measured in long-baseline experiments \cite{Denton:2023zwa}, but JUNO will provide high precision measurements of them \cite{JUNO:2022mxj} and long-baseline experiments will probe the rest.}, test new physics scenarios in oscillations, and address any potential anomalies that arise in upcoming DUNE and HK measurements.
Several experiments have been proposed with varying levels of detail.
One is using a water-based liquid scintillator detector THEIA which combines the advantages of water and liquid scintillator, likely in place of one of the four LArTPC modules for DUNE \cite{Theia:2019non,Theia:2022uyh}.
Another is an additional HK-like tank in Korea experiencing a different portion of the same beam that HK experiences to target the second oscillation maximum in appearance \cite{Hyper-Kamiokande:2016srs}.
There is also discussion about building a very large (much larger than HK) water Cherenkov detector in Sweden combined with additional beam upgrades of the ESS to create the ESSnuSB also to target the second oscillation maximum in \cite{ESSnuSB:2021azq,Alekou:2022mav,Abele:2022iml}.

In this article we discuss a different option, a neutrino factory.
Neutrino factories were seriously considered as an option to probe CP violation in the first decade of this millennium largely due to the expectation of small $\theta_{13}<1^\circ$ which severely limits the possible size of the CP signal \cite{DeRujula:1998umv}.
In general, a neutrino factory is a muon storage ring in a racetrack configuration with long straight sections pointing towards the far detector and provides powerful beams of both muon and electron neutrinos of opposite charges \cite{Geer:1997iz,Blondel:2000gj,Blondel:2006su,FernandezMartinez:2010zza}.
The initial spectrum of this decay is extremely well characterized unlike the initial spectrum from traditional fixed target neutrino oscillation experiments which primarily produce neutrinos from light meson decays with a range of energies.
See \cite{Bogacz:2022xsj} for a recent review of the history of neutrino factories as well as a discussion of the non-oscillation physics program at such a machine.

A neutrino factory may also make sense as a first step on the path to a multi-TeV high energy muon collider to probe the electroweak sector, see e.g.~\cite{Accettura:2023ked}.
This first step could be chronologically as a part of a technology demonstrator, a part of the accelerator chain, or both.
In addition, the recent 2023 P5 report suggested consideration of neutrino physics to be done with a muon accelerator \cite{p5_2023}.

In this paper we will discuss the 
oscillation physics at a neutrino factory considering the current knowledge of the oscillation parameters
and address how a modern look at the oscillation physics case for a neutrino factory compares to previous studies. We then derive numerical sensitivities for the precision on $\delta$, $\theta_{23},\theta_{13}$ and $\Delta m_{31}^2$ for a potential setup of a neutrino factory for two different baselines and energies, before we conclude.

\section{Neutrino Factory Setup}
\begin{figure}
\centering
\includegraphics[width=0.6\columnwidth]{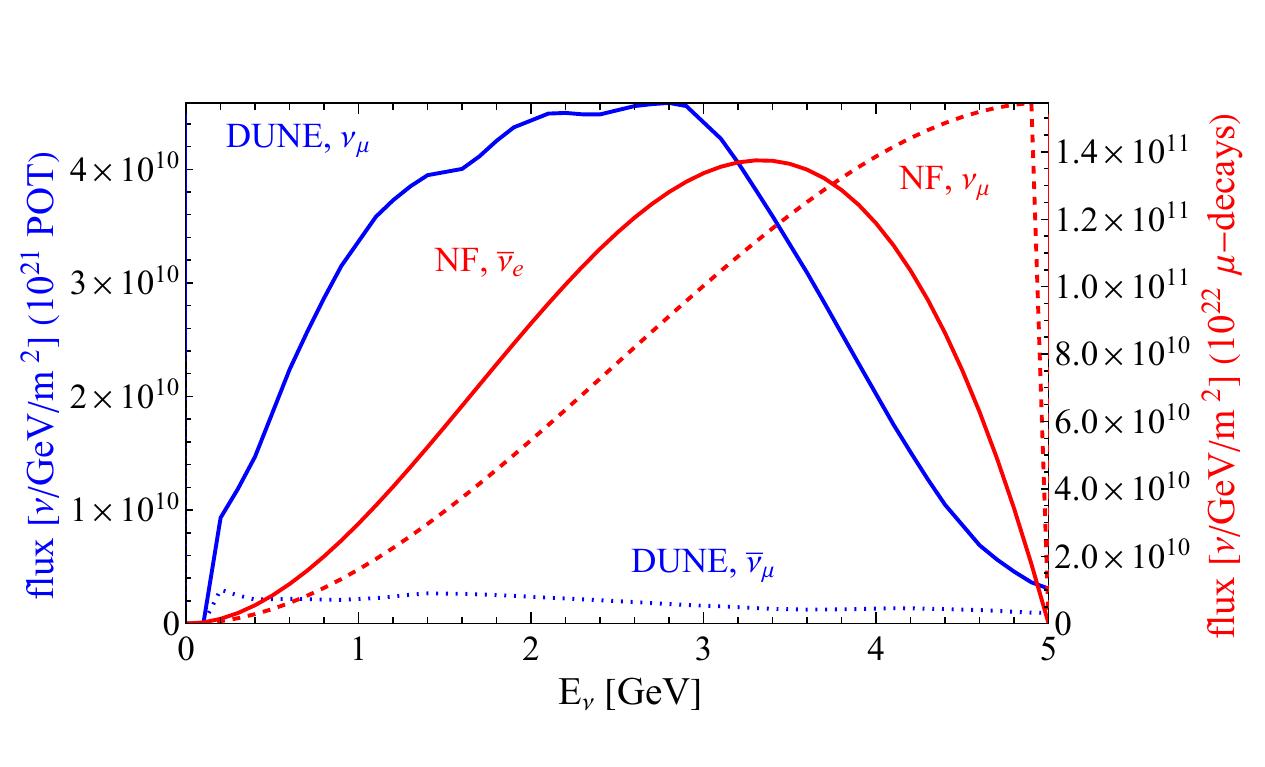}
\caption{The initial neutrino flux composition at DUNE and a neutrino factory from protons on a fixed target (left axis) or muon decays at the energy optimized for the FNAL-SURF baseline: $E_\mu=5$ GeV (right axis).}
\label{fig:fluxes}
\end{figure}

At a neutrino factory the neutrinos originate from the decays of positively or negatively charged muons, leading to a beam of (anti-)electron neutrinos and (anti-)muon neutrinos.
There are a few main differences for the neutrino beam of a neutrino factory compared to conventional neutrino beams from fixed target experiment:
\begin{itemize}
\item The neutrino energies likely reach higher energies at a neutrino factory than a fixed target and the spectrum is different with part of it rising to the maximum, see fig.~\ref{fig:fluxes}.
\item The composition and the expected energy of the neutrino beam is extremely well characterized and radiative corrections to muon decay are precisely known and small \cite{Greub:1993kg,Laing:2008zzb,Tomalak:2021lif}.
Furthermore, no $\nu_\tau$ get produced in the source, allowing for cleaner searches for $\nu_\tau$ appearance channels.
\item The neutrino beam at a neutrino factory is composed of both $\nu_\mu$ and $\nu_e$ and of different charges allowing studies for channels not accessible in other controlled environments including $\nu_e\to\nu_\mu$ and $\nu_e\to\nu_e$; see \cite{Denton:2018cpu}.
\end{itemize}

The first studies of a neutrino factory were focused on measuring the complex phase $\delta$ under the assumption of a very small $\theta_{13}$.
These studies identified numerous interesting approximate degeneracies in the oscillation analysis \cite{Barger:2001yr,Freund:2001ui,Huber:2006wb}. 
For small values of $\theta_{13}$, charge ID (CID) of muons or electrons was crucial to suppress the beam background for the appearance channels and to measure $\delta$.
For this reason, many studies assumed a 100 kT magnetized iron detector \cite{IDS-NF:2011swj}.

With the current and envisioned knowledge of all oscillation parameters and the present experimental landscape, the task of a future neutrino factory is no longer to discover the oscillation parameters, but rather to make precise measurements of the oscillation parameter and potentially resolve any discrepancies identified in previous measurements.
Therefore, with a different physics goal and a more modern experimental landscape, the experimental requirements on a neutrino factory also necessarily evolve.

We consider two possible neutrino factory scenarios:
the neutrino source is Fermilab and, to make use of existing facilities, the far detector to be located at SURF, resulting in a baseline of 1284.9 km. Alternatively, we also consider the neutrino source at Brookhaven National Laboratory (BNL), to make use of the AGS/RHIC/EIC accelerator complex, which leads to baseline of 2542.3 km to SURF.\footnote{See \cite{Kitano:2024kdv} for the study of a neutrino factory based at J-PARC.}
We assume as matter density for both configurations 2.848 g/$\text{cm}^3$; the average matter density for the BNL configuration would actually be somewhat higher which would slightly enhance the differences between the two seen below.
We assume that the far detector is a LArTPC configuration with a total fiducial target mass of 40 kT consistent with a possible final version of the envisioned DUNE detector.
Based on previous exploratory studies of neutrino factories we assume $10^{21}$ muon (or anti-muon) decays per year \cite{IDS-NF:2011swj}.
This would require a comparable (to within $\mathcal O(1)$ factors) beam power as the 1.2 MW beam expected for DUNE.

To simulate a neutrino factory we make use of DUNE's far detector design \cite{DUNE:2021cuw} in the GLoBES \cite{Huber:2004ka} framework.
We also consider oscillated tau neutrinos events with a tau neutrino CC cross section from \cite{psf2023008064}, although we treat the $\nu_\tau$ events as background only \cite{Donini:2010xk} and don't consider tau signal events for this three-flavor oscillation study due to the large uncertainties on the signal efficiency at LAr detectors.
We have verified that including this channel would only marginally improve the sensitivity to measure the oscillation parameters because the $\nu_\tau$ cross section is significantly suppressed at these energies.

While we do not perform a full end-to-end simulation of DUNE or a neutrino factory as that is beyond the scope of this work and would require many detector and accelerator details that are still under development, we do implement uncertainties designed to match what is generally available.
To simulate DUNE, we use the files provided by DUNE in \cite{DUNE:2021cuw}, which largely come from \cite{DUNE:2015lol} and slightly modify them.
For DUNE, we take the conservative estimates for the muon and electron neutrino event rate normalization uncertainties of 5\% and 2\%, respectively.
The muon neutrino uncertainty taken to be larger than the electron neutrino uncertainty because it includes determinations of the beam flux, many cross section parameters, and the fiducial volume, consistent with the approach provided by DUNE.
The electron neutrino uncertainty does include a contribution due to the translation of those measurements into the appropriate values for electron neutrinos.
For the beam power we take 1.2 MW with 56\% uptime as recommended\footnote{Note that for the neutrino factory we consider only the total number of stored muons per year, which thus have any uptime effects already accounted for.}.
While this run time configuration for DUNE of 480 kT-MW-years is not exactly identical to the latest DUNE staging, it is quite similar to the expected exposure after 10 years of DUNE \cite{DUNE:2022aul,DUNE:2024wvj} and provides for an important benchmark to compare with a neutrino factory.

For a neutrino factory going to the same far detector, we take the same approach with some improvements.
We remove the uncertainty due to the detector volume as we only consider neutrino factory statistical tests simultaneously with DUNE measurements, so the same uncertainty should not be double counted.
We also remove the flux uncertainty as it will be significantly reduced due to the well characterized beam from muon decay.
Finally, we split the cross section model uncertainty equally across both flavors to find 2.5\% uncertainty for both flavors because the statistics in the FD is comparable between both flavors.

For the backgrounds we keep with the recommendations of \cite{DUNE:2021cuw} and 10\% on the normalization of the NC background, 20\% on the normalization of the tau background\footnote{See e.g.~\cite{DeGouvea:2019kea,psf2023008064} for similar numbers.
We are slightly more optimistic (25\%$\to$20\%) as this number includes a flux uncertainty measured at the near detector which should be improved at a neutrino factory.}, and 2\% on the Earth's density.

\section{Sensitivities}
For our simulations we consider priors on the oscillation parameters other than $\delta$ from \cite{Gonzalez-Garcia:2021dve} and we fix the MO to the normal ordering (NO), as the MO as well as all oscillation parameters will be determined via multiple independent methods with future experiments. For comparison we also show the results for true inverted ordering IO in  appendix \ref{sec:BNLres}.
We focus our sensitivity studies on $\delta$ as this is the least well-measured parameter in neutrino oscillations, but we also investigate a neutrino factory's sensitivity to the other relevant oscillation parameters: $\Delta m^2_{31}$, $\theta_{23}$, and $\theta_{13}$, while long baseline experiments only have limited sensitivity to the solar parameters \cite{Denton:2023zwa}.
The precision on $\delta$ does provide our primary metric quantifying the ability of a neutrino factory to differentiate among differing measurements from upcoming experiments.

Since the beam at a neutrino factory consists of equal amounts of neutrinos and anti-neutrinos, there is an inherent large contribution to the final states coming from the other charge, i.e.~the processes $\nu_e\to \nu_\mu$ and $\bar{\nu}_\mu\to \bar{\nu}_\mu$ lead to the same final state although with a different energy spectrum due to different initial spectra and different oscillation probabilities, see appendix \ref{sec:events}. Differentiating the events from the disappearance channel in the analysis of the appearance channel has been identified as an important step towards a successful neutrino factory early on \cite{DeRujula:1998umv}. Therefore we study the impact of CID efficiency $\epsilon_{CID}$, which we take to be energy independent, in terms of either electron or muon charge identification.
We define the CID parameter $\epsilon_{CID}$ as
\begin{align}
N_{\nu_\text{f,obs}}&=\frac{\epsilon_f}2\left[(1+\epsilon_{CID}) N_{\nu_f}+(1-\epsilon_{CID})N_{\bar\nu_f}\right]\\
N_{\bar\nu_\text{f,obs}}&=\frac{\epsilon_f}2\left[(1+\epsilon_{CID}) N_{\bar\nu_f}+(1-\epsilon_{CID})N_{\nu_f}\right]
\end{align}
with the detection efficiency $\epsilon_f$ and $N_\nu$ is the number of (anti-)neutrinos of flavor $f$ we get after taking oscillations of the initial flux into account.
DUNE could potentially differentiate a muon final state from an anti-muon final state with $\epsilon_{CID}\approx 72\%$ \cite{Ternes:2019sak} from muon capture on argon. In addition, inelasticity can provide some CID information as well.

Larger values of $\epsilon_{CID}$ could be achieved with magnetized detectors, an idea pursued by the INO collaboration. They estimate a CID efficiency of $\epsilon_{CID}\gtrsim 95\%$ for muons with GeV energy \cite{ICAL:2015stm}. Electron CID at GeV energies has been studied in  the scenario of a totally active scintillator detector in a magnetic field \cite{Bross:2007ts,Asfandiyarov:2014haa}, and a magnetized LAr detector \cite{Rubbia:1977zz,Rubbia:2001pk,Rubbia:2004tz,Rubbia:2009md}.
Given that the final DUNE detector modules remain undesigned, a discussion of the possible benefit of working to improve CID is quite timely. 
In \cite{Huber:2008yx} it has been pointed out that CID can also be achieved statistically at some level, provided that the energy resolution is sufficiently good.

To compare our results for a NF 
and to get reliable conclusions about the potential state of oscillation physics with an inclusion of a NF after the next generation of LBL experiments we combine and compare  the NF results to DUNE's and HK's forecasted results.
For DUNE we assume 480 kT-MW-year which corresponds to 5 years of each neutrino running and anti-neutrino with 1.2 MW proton beam and with a total fiducial volume of 40 kT of LAr.
For HK we assume 190 kT water detector, 1.3 MW beam running for 10 years with $\nu:\bar{\nu}=1:3$ from \cite{Scott:2020gng}. We simulate both experiments within the GLoBES framework.

\begin{figure}
\centering
\includegraphics[width=0.6\columnwidth]{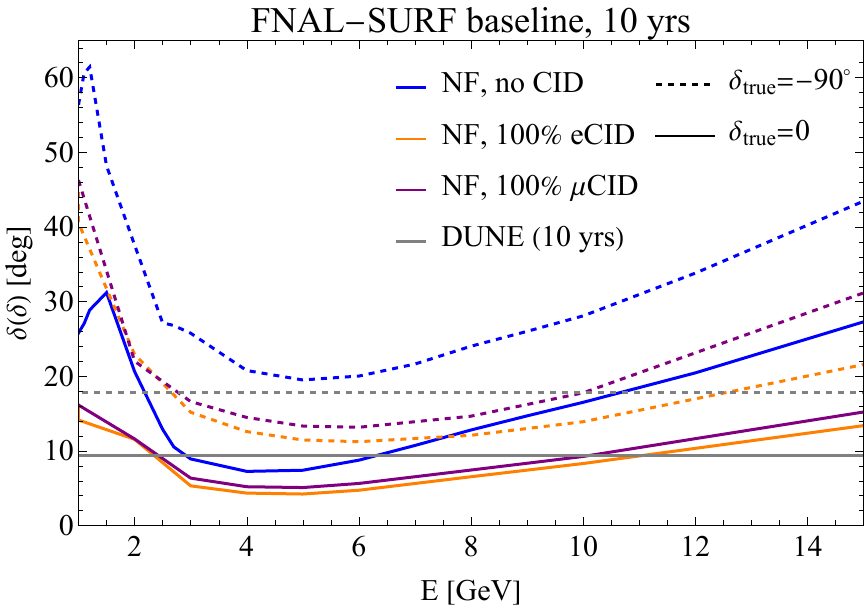}
\includegraphics[width=0.6\columnwidth]{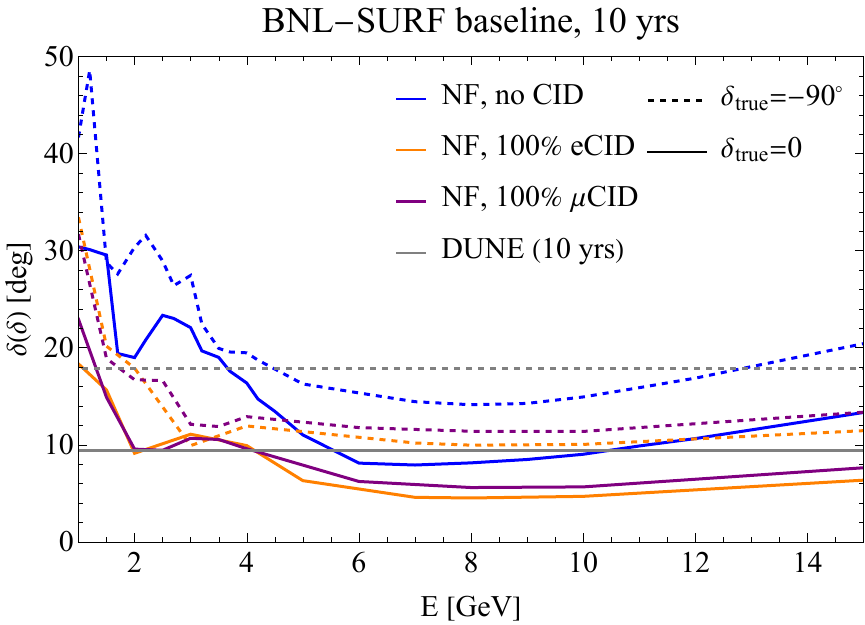}
 \caption{The expected $1\sigma$ precision on $\delta$ as a function of the muon energy for a total of 10 years at a NF (without inputs from DUNE/HK) assuming $10^{21}$ muon decays per year, for two different baselines and two different true values of $\delta$ in true NO. The colors show different assumption of CID capabilities. For comparison, the horizontal lines indicate the expected sensitivity of 10 years of DUNE.}
\label{fig:muon energy}
\end{figure}

\begin{figure}
\centering
\includegraphics[width=0.6\columnwidth]{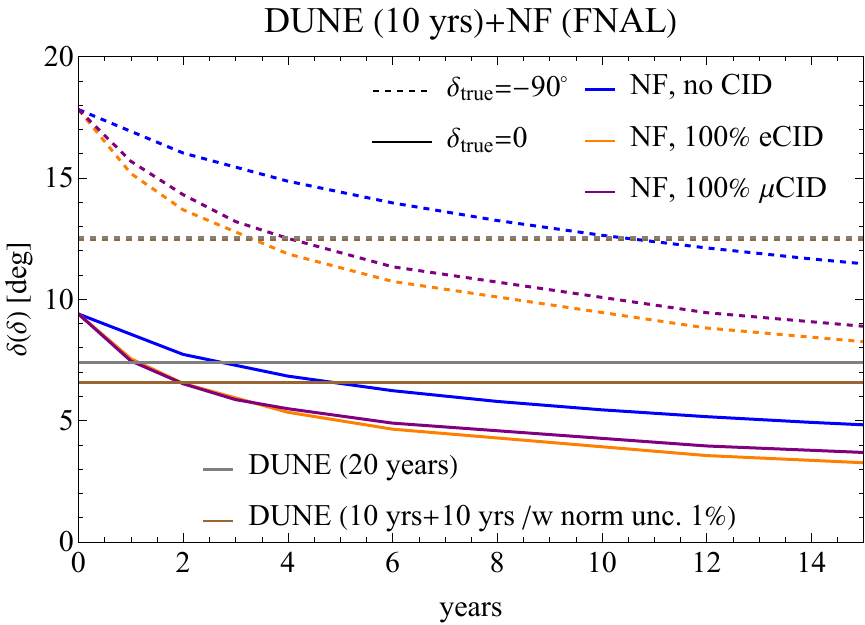}
\includegraphics[width=0.6\columnwidth]{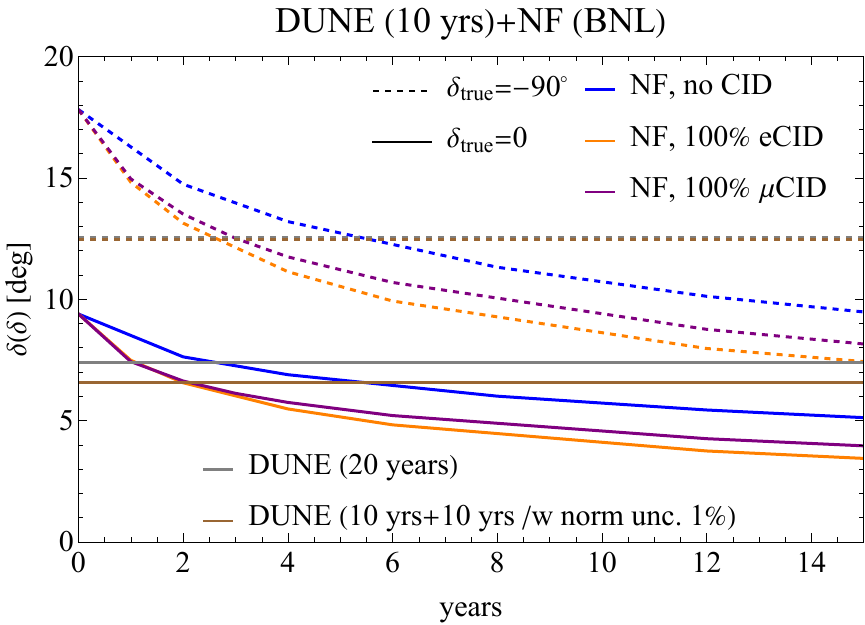}
\caption{The expected $1\sigma$ precision on $\delta$ as a function of the running time of a NF for two different baselines and two different true values of $\delta$ and NF-FNAL energy of 5 GeV, NF-BNL energy of 8 GeV. We assume $10^{21}$ muon decays per year. We assume that the NF is combined with 10 years of DUNE. For comparison we show the expected sensitivity for 20 years of DUNE with the nominal systematic uncertainties, or assuming that electron and muon neutrino normalization uncertainties will be reduced to 1\% in the final 10 years.
}
\label{fig:NFtime}
\end{figure}

\subsection{Sensitivity to $\delta$}
In fig.~\ref{fig:muon energy} we show the 1$\sigma$ sensitivity to $\delta^\text{true}=(-90^\circ,~0)$ as a function of muon energy for a total of $10^{22}$ muon decays and a 40 kT FD in true NO.
For the FNAL-SURF baseline we find that a muon energy of $E_\mu\simeq5$ GeV with equal muon and anti-muon running time allows for the most precision on the determination of $\delta$ for most cases.
For the BNL-SURF baseline we find a higher energy $E_\mu\simeq8$ GeV is optimal, although higher energies have similar results.
The higher energy is necessary for the longer baseline to get the appearance maximum to occur within the muon decay spectrum at that baseline. In general we find that the optimal region covers a  range of energies as the neutrino flux  at a NF has a broad spectrum. Assuming perfect CID abilities slightly increases the optimal energy range. For these optimal energies a NF can improve over 480 kT-MW-year of DUNE assuming $10^{22}$ total muon decays.
An even lower muon energies a NF could be sensitive to the second oscillation maximum as demonstrated by the dips in the precision curve in fig.~\ref{fig:muon energy}.

Depending on the specific details of the experimental configuration as well as progress in measuring the $\delta$ and the other oscillation parameters may lead to slight modifications on these optimal energies. In the following we will fix the optimal muon energies to 5 GeV for the FNAL setup and 8 GeV for the BNL setup.\footnote{In app.~\ref{sec:BNLres} we show the results for higher energies for the BNL setup as well.} 

In fig.~\ref{fig:NFtime}
 we show the  $1\sigma$ precision on $\delta$ for $\delta^\text{true}=(-90^\circ,~0)$ for the two setups for the optimal muon energies of 5 Gev and 8 GeV as a function of the total number of muon decays. We assume that the NF is combined with 480 kT-MW-year of DUNE. To  improve the precision on $\delta$ independent of the true value of $\delta$ compared to  960 kT-MW-year  of DUNE a total of $\gtrsim 11\times 10^{21}~\mu$ decays is required for the FNAL setup without CID but only $\gtrsim4\times 10^{21}~\mu$ if CID can be realized. For the BNL setup even less muon decays are required, $\gtrsim 6\times 10^{21}$ decays without CID, $\gtrsim 3\times 10^{21}$
 with CID. Even if the normalization uncertainty on the DUNE flux can be decreased to 1\%, a NF can still lead to improvements on the precision over 480 kT-MW-year of DUNE with nominal uncertainties combined with 480 kT-MW-year with reduced uncertainties.
 In the following we fix the total number of muons to $10^{22}$ for a 40 kT fiducial mass FD. This could correspond to running of 10 years, equally divided between muons and anti-muons, with $10^{21}$ (anti-)muon per year.
 
\begin{figure*}
\centering
\includegraphics[width=0.49\columnwidth]{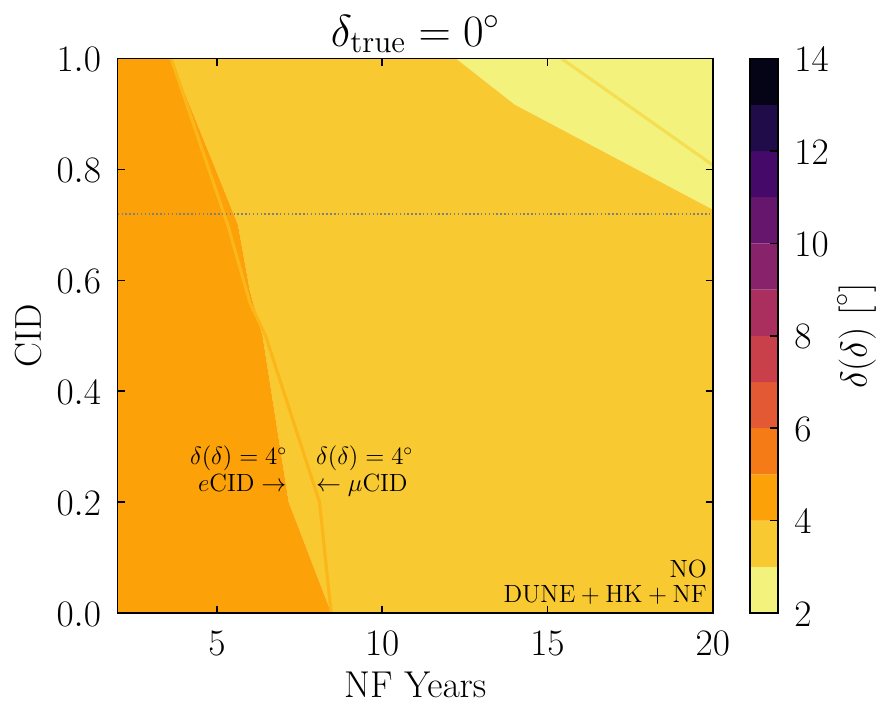}
\includegraphics[width=0.49\columnwidth]{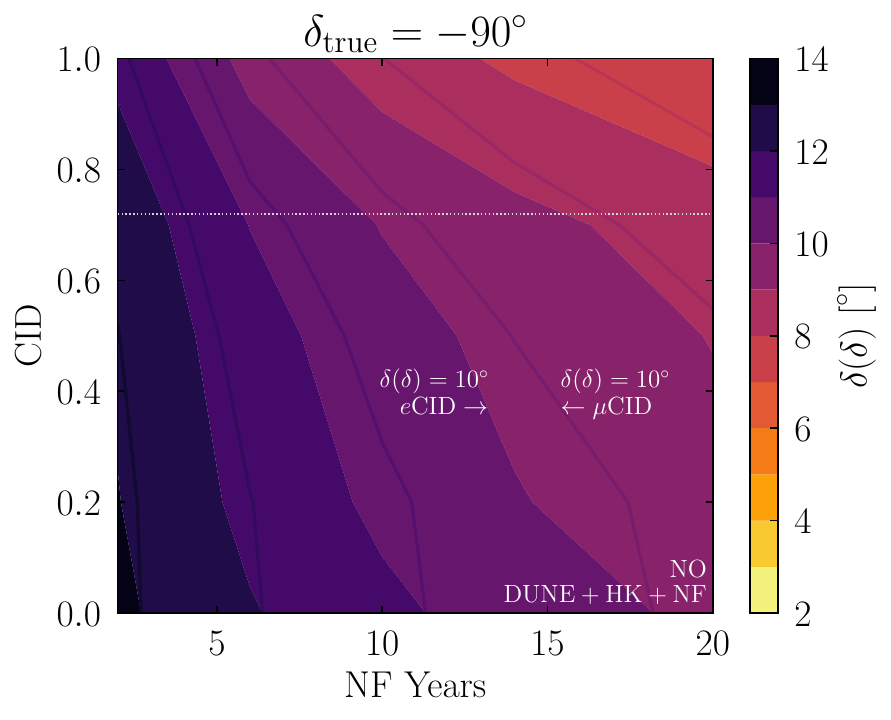}
\caption{Precision on $\delta$ in degrees as a function of neutrino factory exposure and CID.
The neutrino factory has a baseline and muon energy of 1284.9 km and 5 GeV and also includes 10 years of both DUNE and HK.
The shaded regions are for electron CID and the lines are for muon CID; electron CID does slightly better in most cases.
The horizontal lines show possibly achievable muon CID.
The true values of $\delta$ are $0$ (\textbf{left}) and $-90^\circ$ (\textbf{right}).
We assume as ratio of neutrino to anti-neutrino running time of 1:1 for the neutrino factory and DUNE 1:1 and 1:3 for HK.  We assume true NO.}
\label{fig:precision_delta}
\end{figure*}

To show the effect of CID in more detail we show in fig.~\ref{fig:precision_delta}  the expected $1\sigma$ precision on $\delta$ as a function of CID and exposure for the FNAL-SURF setup for $\delta_{\text{true}}=0,-90^\circ$.
We include the anticipated information from 10 years each of DUNE and HK.
Our results show that CID can improve the precision on $\delta$ by 15-20\%.
We have also studied the effect of assuming muon and electron CID at the same time and the results are similar. 
We find that electron CID has a bigger impact on the improvements in precision on $\delta$ than muon CID.
This is due to the fact that the detected $\nu_\mu\to \nu_e$ spectrum is more similar to the detected $\bar{\nu}_e\to \bar{\nu}_e$ spectrum than the detected $\nu_e\to \nu_\mu$ spectrum to the detected $\bar{\nu}_\mu\to \bar{\nu}_\mu$ spectrum, see appendix \ref{sec:events}.

For the BNL baseline the improvement on the precision on $\delta$ does not significantly changes  between  6-14 GeV, in particular if we assume CID. Constructing a neutrino factory with a higher muon energy could be preferential to avoid too many fast muon decays  and for R\&D studies of muon cooling technologies. Therefore we also study the precision on the oscillation parameters at these higher energies. The results can be found in the appendix \ref{sec:BNLres} in tab.~\ref{tab:resultshigherE}. 

We find that the sensitivity of a neutrino factory comes somewhat more from the $\nu_e\to \nu_\mu$ golden channel \cite{DeRujula:1998umv,Cervera:2000kp} and less from the $\nu_\mu\to \nu_e$ channel which will be leveraged by future proton accelerator neutrino experiments.
The oscillation probabilities in these two channels are the same up to the change $\delta\to-\delta$, so each channel serves as an important cross check of the other, but provides no fundamentally new insights into the nature of CP violation or conservation.
They do, however, provide a probe of CPT invariance because a neutrino factory is sensitive to four different appearance channels which CP, T, and CPT conjugates of each other.
Thus a neutrino factory, especially when combined with DUNE and HK, will provide a valuable unique test of CPT invariance.

Overall, the improvements on precision on $\delta$ with the BNL-SURF setup are larger than at the FNAL-SURF setup largely because in the former setup the neutrinos experience more matter effects \cite{Wolfenstein:1977ue} due to the higher neutrino energy which increases the appearance probability. Furthermore, due to the higher neutrino energy in this setup the cross section uncertainty is smaller.
Finally, even though the flux is lower in the BNL-SURF setup due to the larger baseline and beam spreading, the cross section is higher, the appearance probability is somewhat larger, and the variation in the probability due to variations in $\delta$ is also somewhat larger; these effects provide some cancellation for the lower flux.

\begin{figure}
\centering
\includegraphics[width=0.6\columnwidth]{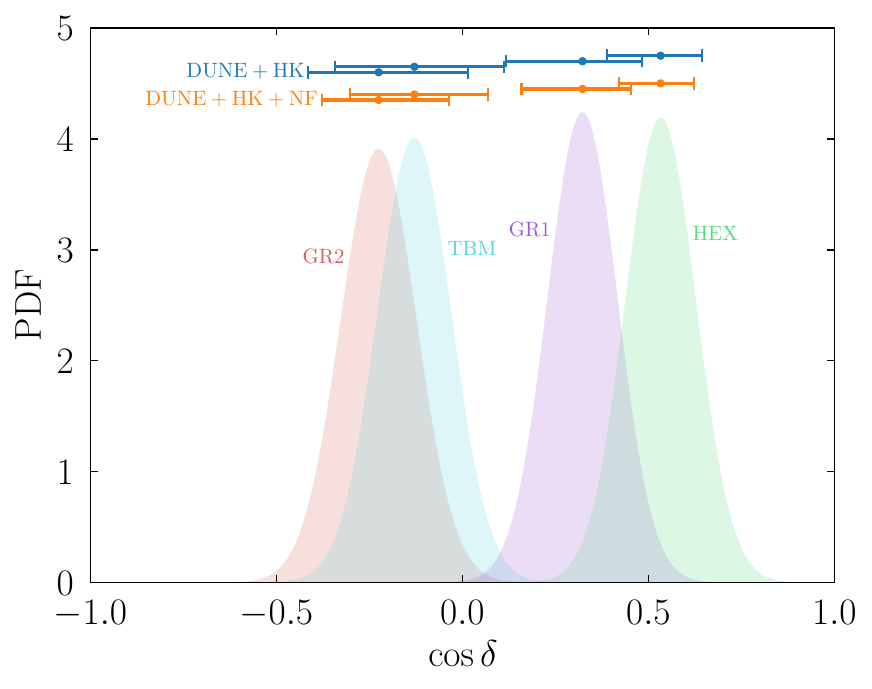}
\caption{The predicted regions of $\cos\delta$ due to several popular structures within the charged-lepton correction framework, adapted from \cite{Everett:2019idp}; GR1 and GR2 are two flavor model predictions based on the golden ratio \cite{Datta:2003qg,Everett:2008et,Rodejohann:2008ir,Adulpravitchai:2009bg,Feruglio:2011qq,Ding:2011cm}, TBM is the tribimaximal flavor model prediction \cite{Wolfenstein:1978uw,Harrison:2002er,Harrison:2002kp,Xing:2006xa,He:2003rm}, and HEX is based on a hexagonal structure \cite{Albright:2010ap}.
Along the top are the expected $1\sigma$ sensitivities to $\cos\delta$ at the various central values predicted by the models.
DUNE and HK are both at 5+5 years as described in the text, as is neutrino factory with the FNAL baseline and no CID.}
\label{fig:flavor}
\end{figure}

While increased precision on $\delta$ and other oscillation parameters for the sake of precision is an important goal of particle physics, we also provide some context for these sensitivities.
Because $\delta$ is the last remaining oscillation parameter that is largely unconstrained, it plays a key role in flavor model predictions \cite{Girardi:2014faa,Everett:2019idp}, along with other parameters such as $\theta_{12}$ and $\Delta m^2_{21}$ \cite{Gehrlein:2022nss} which are expected to become measured with much more precision in the coming years by JUNO \cite{JUNO:2022mxj}.
While there are many conceivable predictions on neutrino parameters from many different structures (see \cite{Denton:2023hkx} for a recent survey), here we focus on charged lepton corrections as they predict the mixing parameters and not the masses.
Charged lepton corrections start with a mass matrix for neutrinos derived from some symmetry leading to popular and well-known structures such as tri-bimaximal (TBM) which predicts $\theta_{12}\approx35^\circ$, $\theta_{23}=45^\circ$, and $\theta_{13}=0$, and thus no CP violation.
A small correction is thus needed to explain $\theta_{13}=8.5^\circ$ which also allows for CP violation, a slightly smaller value of $\theta_{12}$, and possible variation in $\theta_{23}$; these are usually provided in the charged lepton mass matrix which, in addition to the neutrino mass matrix, provides a slightly off-diagonal matrix parameterized by a Cabibbo sized angle.
In fig.~\ref{fig:flavor} we show the predictions of $\cos\delta$ from four popular charged lepton correction structures as discussed in \cite{Everett:2019idp} alongside the expected $1\sigma$ precision on $\cos\delta$ (taking the wider allowed ranges in all cases due to the sign degeneracy since long-baseline experiments predominantly measure $\sin\delta$ from their appearance channels, although they also probe $\cos\delta$ from their disappearance measurements \cite{Denton:2023qmd}).
We consider both the precision from the combination of DUNE and HK as well as the expected improvement from including the neutrino factory at the FNAL baseline.
In table \ref{tab:results} we document the expected precision on $\delta$ for various experiments.

We see that while after DUNE and HK there will be some modest model discriminating capabilities in some cases, the addition of neutrino factory will somewhat increase the capability to differentiate flavor models in some standard benchmark cases as shown in fig.~\ref{fig:flavor}.
Significant additional improvement in model discrimination capability is likely to also come from JUNO's expected improved determination of the solar parameters in coming years as well.

\subsection{Sensitivity to other oscillation parameters}
\begin{figure}
\centering
\includegraphics[width=0.45\columnwidth]{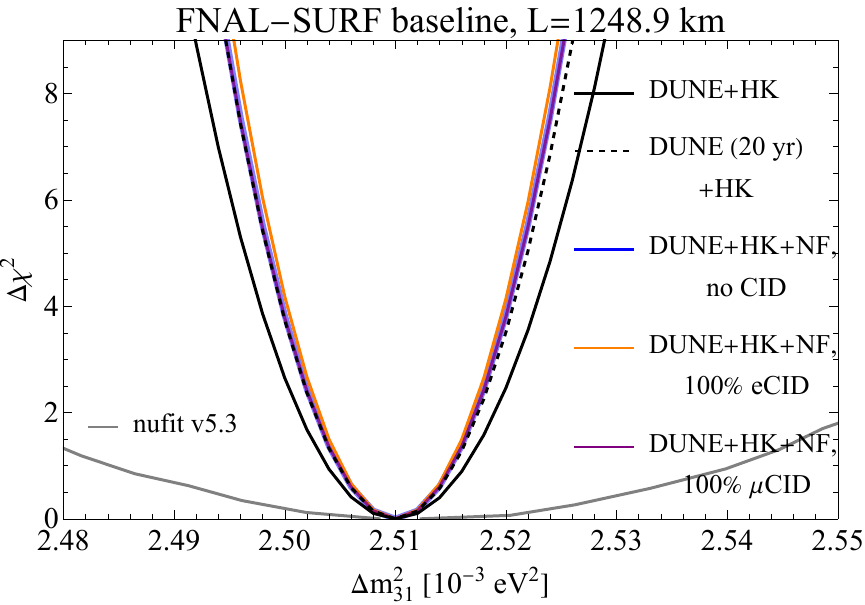}
\includegraphics[width=0.45\columnwidth]{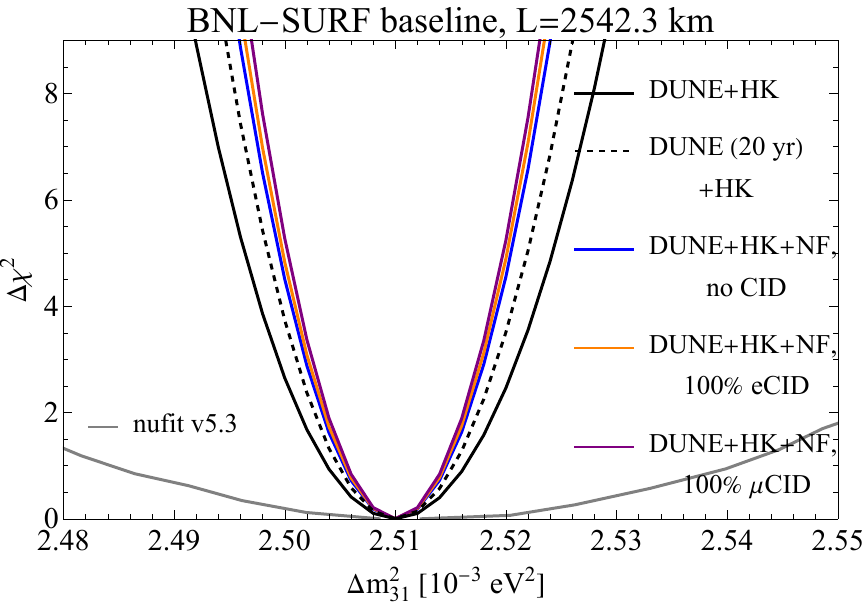}
\includegraphics[width=0.45\columnwidth]{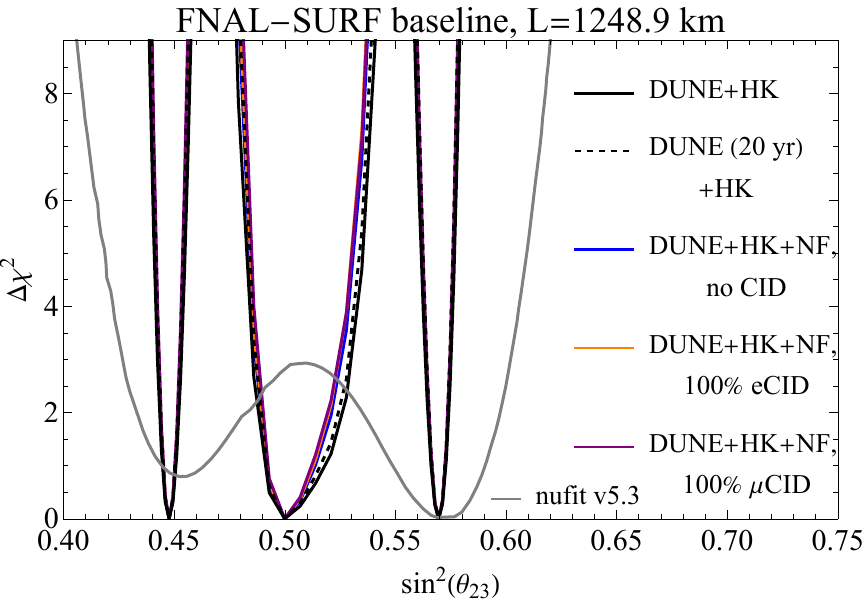}
\includegraphics[width=0.45\columnwidth]{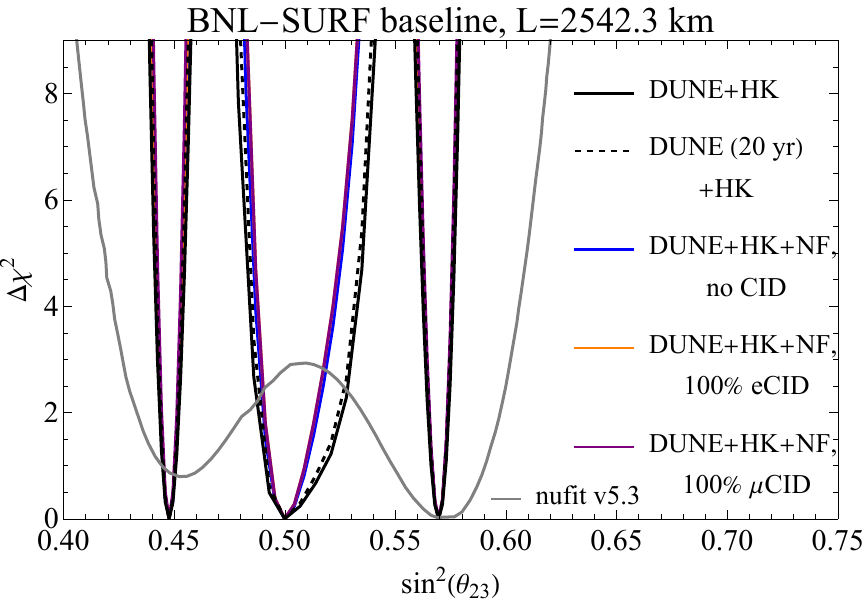}
\includegraphics[width=0.45\columnwidth]{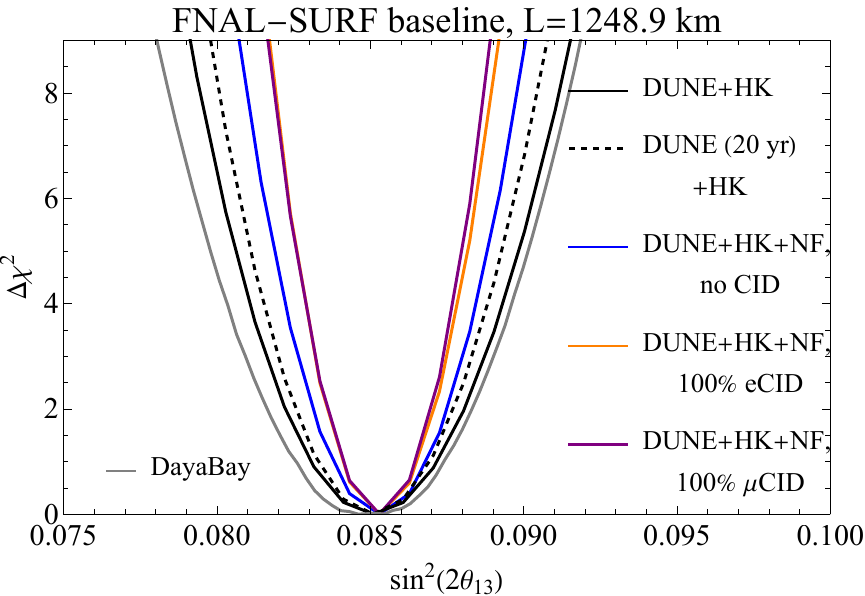}
\includegraphics[width=0.45\columnwidth]{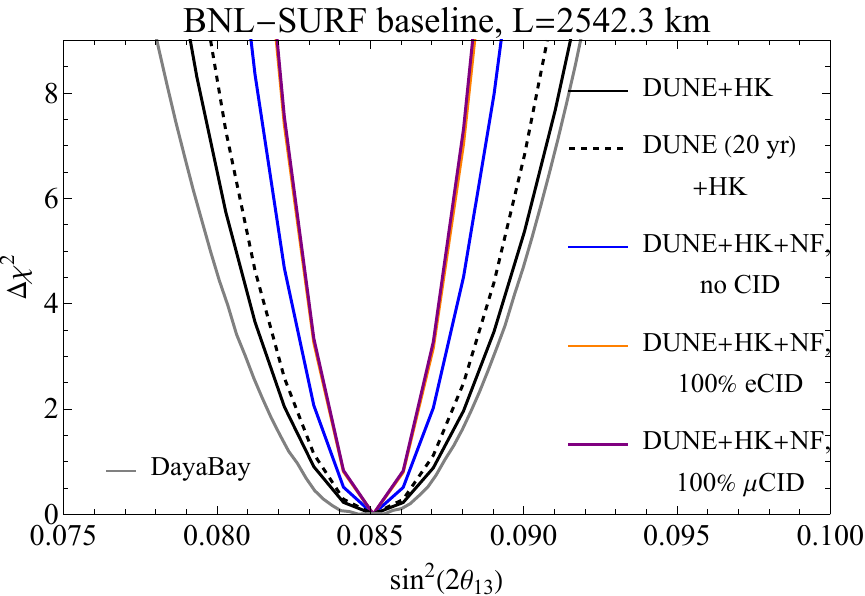}
\caption{Precision on $\Delta m_{31}^2, ~\sin^2(2\theta_{13}),~\sin^2 \theta_{23}$ for true NO. For $\theta_{23}$ we assume three different true values $\theta_{23}^\text{true}=(42^\circ,~45^\circ,~49^\circ)$. We show the results for 10 years of NF with $10^{21}$ muon decays per year and a 40 kT FD and $E_\mu=5 ~(8)$ GeV for the FNAL (BNL) setup on the left (right) plots, and for different assumptions on CID. The results for the NF also include 10 years of both HK and DUNE. 
We also compare the results to 20 years of DUNE in addition to 10 years of HK. The gray curves show the current precision on these parameters from \cite{Esteban:2020cvm,DayaBay:2022orm}.}
\label{fig:otherparameters}
\end{figure}

\begin{figure}
\centering
\includegraphics[width=0.47\columnwidth]{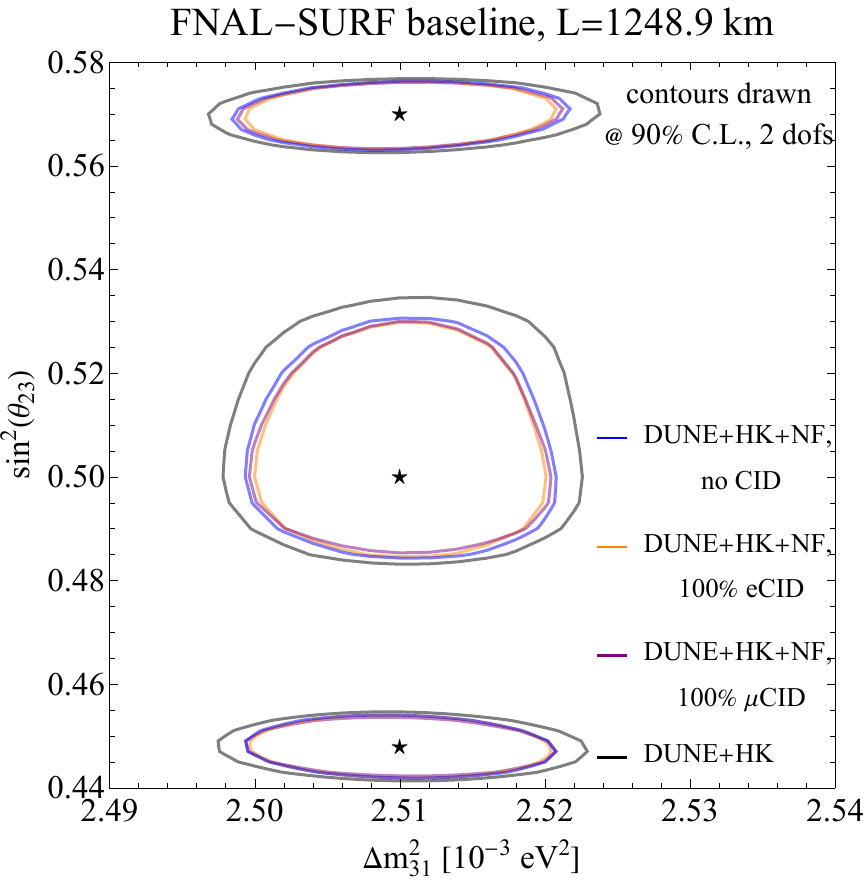}
\includegraphics[width=0.45\columnwidth]{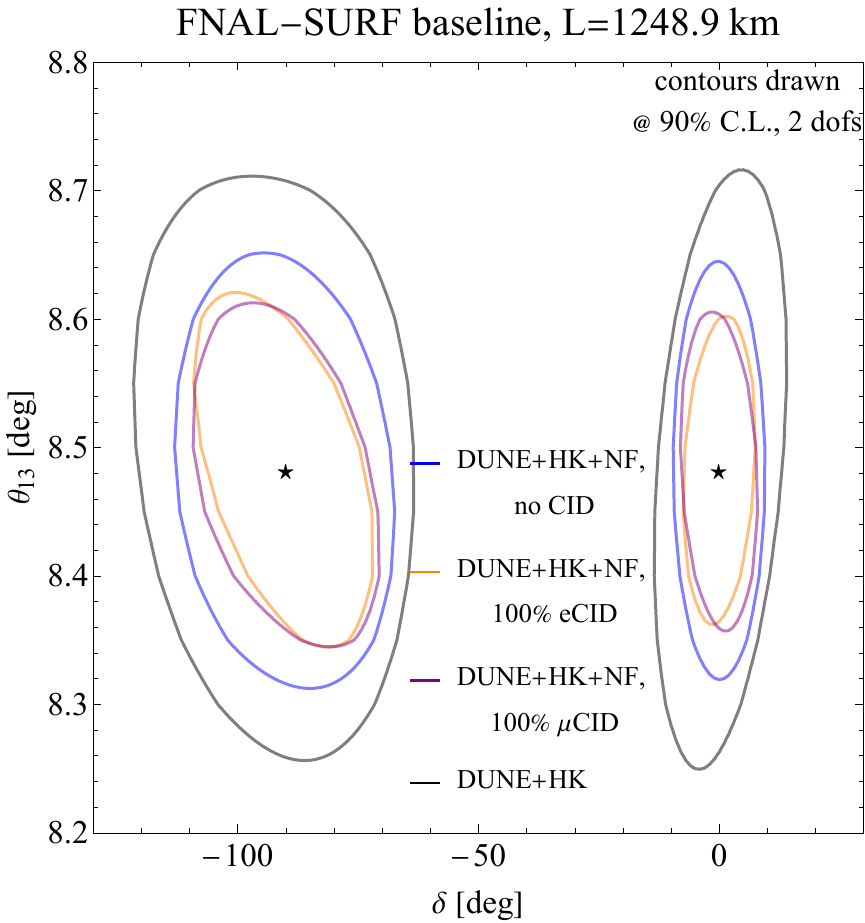}
\caption{Two-dimensional precision on the oscillation parameters assuming DUNE+HK+NF (10 years of each) with the FNAL-SURF baseline assuming different true values marked with a star. The contours are drawn at 90\% C.L. for 2 dofs. We assume true NO. }
\label{fig:otherparameters2d}
\end{figure}

In fig.~\ref{fig:otherparameters} we show the  precision on  $\sin^2\theta_{23},~\sin^22\theta_{13}$ and $\Delta m_{31}^2$\footnote{Just like DUNE and HK a NF has only very limited sensitivites to the solar parameters \cite{Denton:2023zwa}.} assuming 10 years of a NF with $10^{21}$  muon decays per year combined with 10 years of  DUNE and HK each.  We compare the results to existing constraints from global analyses and Daya Bay, and the expected sensitivies from DUNE and HK alone, and 20 years of DUNE  and 10 years of HK. We summarize the results in tab.~\ref{tab:results_others}.
In fig.~\ref{fig:otherparameters2d} we show the correlations between $\sin^2\theta_{23}-\Delta m_{31}^2$ and $\theta_{13}-\delta$.

We find that the combination of DUNE and HK will drastically improve the precision on the atmospheric parameter $\Delta m_{31}^2$ and $\theta_{23}$.
The addition of a NF will further improve the precision nearly independent on the assumption of CID.
We focus on cases where the octant of $\theta_{23}$ can easily be identified, but a NF will also increase the sensitivity to determining to correct octant for regions of parameter space close to maximal mixing.
The improvement on the precision on $\Delta m_{31}^2$ is slightly larger for the BNL setup where it can also improve over 20 years  of DUNE combined with 10 years of HK. For $\theta_{23}$ the improvements in precision due to the addition of a NF are most noticeable for $\sin^2\theta_{23}^\text{true}=0.5$ where a NF leads to a more symmetric precision. Again we find that the BNL setup leads to slightly higher precision that the FNAL setup and can even improve over 20 years  of DUNE plus 10 years of HK.
Finally, DUNE and HK lead to slight improvement on the precision of $\sin^2 2\theta_{13}$ over Daya Bay which can even be further increased by the addition of a NF. Both setups could even improve over 20 years  of DUNE and 10 years of HK with additional improvements coming from CID of either electrons or muons.
Also the two-dimensional precision on $\Delta m_{31}^2-\sin^2\theta_{23}$ and $\theta_{13}-\delta$ benefit from the addition of a NF to DUNE and HK.

Going to higher muon energies for the BNL setup  slightly worsens the oscillation parameters but still provides an improvement over the combination of just DUNE and HK, for some oscillation parameters even compared an increased exposure for DUNE, see tab.~\ref{tab:resultshigherE} in app.~\ref{sec:BNLres}.

\section{Discussion}
This study advances beyond others in several important ways.
First, being the first such study in many years, it makes use of improved knowledge of the oscillation parameters; not just $\theta_{13}$, but also $|\Delta m^2_{31}|$ which has steadily improved with measurements from 7+ different experiments.
Second, the next generation global long-baseline experimental picture is shaping up.
There will be a precision reactor experiment (JUNO), two advanced long-baseline accelerator experiments (DUNE and HK), and improvements with atmospheric neutrinos (IceCube, HK, and KM3NeT).
Third, the US has committed to a program of LArTPC's for good energy resolution and particle identification.
Fourth, the US interest in a potential muon collider has recently increased leading to the serious possibility of building a muon accelerator, technology that may well require a lower energy machine to demonstrate muon cooling.
Finally, we have carefully investigated all relevant aspects of three-flavor oscillations including CID, the role of systematics, and the impact to flavor model predictions.

We now highlight the differences between a neutrino factory and DUNE as presented here.
First, the shape of the spectrum is quite different and it tends to rise to the high energy part of the spectrum.
This is compared to the broader spectrum that DUNE will see based on the specific horn configuration designed to have good sensitivity to $\delta$ \cite{DUNE:2020ypp,Calviani:2014cxa}.
Second, the spectrum of neutrinos for a neutrino factory is precisely predicted, while that for DUNE carries significant uncertainties.
Third, a neutrino factory provides access to more channels (notably $\nu_e\to\nu_e$, $\nu_e\to\nu_\mu$, and their conjugates) than are regularly easily accessible.
Fourth, for a given accelerator complex, the flux of neutrinos from a neutrino factory will be similar to, but likely less, than that for DUNE.
The exact details here depend on muon cooling efficiencies and the geometry of the neutrino factory, but all factors in this point are no more than $\mathcal O(1)$ factors.

Our results show that while CID does enhance the oscillation parameters' precision, CID plays a smaller role than emphasized in studies a decade ago.
This is primarily due to the fact that LArTPC detectors have sufficiently good energy resolution to distinguish signal from background as well as identify the different flavors.
In fact, the LArTPC energy resolution could be even better than assumed here, see e.g.~\cite{DeRomeri:2016qwo,Friedland:2018vry,Kopp:2024lch}. 

A neutrino factory would also improve the precision on other oscillation parameters, in particular $\theta_{23}$ and $\Delta m^2_{31}$ which have important connections to flavor models as well as probes of the absolute neutrino mass scale such as cosmology and neutrinoless double beta decay.
Indeed, a neutrino factory could decrease the 1$\sigma$ uncertainty on $\theta_{23}$, in particular for true maximal values of $\theta_{23}$ and smaller improvements for $\theta_{23}^\text{true}=(4.2^\circ,~49^\circ)$. Similarly, the precision on
$\theta_{13}, ~\Delta m_{31}^2$ can be improved as well. Generally, the improvements due to the addition of a NF are more prominent for true NO.
However a neutrino factory has only limited sensitivity to the solar parameters, just like DUNE \cite{Denton:2023zwa}. 

Using a similar detector setup for a neutrino factory as for near-future experiments could further reduce the uncertainties on the cross section and detector efficiencies, as there is an ongoing  program to measure the neutrino cross section on Ar and advancing the LAr technology \cite{Machado:2019oxb}.

A neutrino factory could be an appealing possible future neutrino oscillation experiment should the results of HK and DUNE disagree.
If the weak tension between NOvA and T2K in the CP violation measurements \cite{Denton:2020uda,Chatterjee:2020kkm} persists, DUNE will be able to probe it \cite{Denton:2022pxt}. However if the tension requires further oscillation studies at a higher neutrino energy, longer baseline, and  overall smaller flux uncertainties than fixed target oscillation experiments a neutrino factory is a favorable setup for future studies.\footnote{Note that atmospheric neutrino experiments can also provide oscillation studies at higher energies and longer baselines however their flux uncertainty and flavor composition is more uncertain than at a neutrino factory.} Furthermore, at a neutrino factory the neutrino energy is flexible and tunable \cite{Delahaye:2013jla}. Finally, at a neutrino factory six oscillation channels and their CP conjugate ones are accessible with similarly large number of events as the initial neutrino  beam consists equally of $\nu_e, \nu_\mu$.
The presence of numerous oscillation channels provides multiple independent means of getting at the oscillation parameters, each with different dependences on the systematic uncertainties.

\begin{table}
\centering
\caption{Resolution on $\delta$ for combinations of different experiments for two true values of $\delta$ in NO. For DUNE we assume 480 kT-MW-yr, for HK we assume 2.47 MT-MW-yr, for the neutrino factory exposure we assume  a 40 kT LAr detector and a total of $10^{22}$ muon decays -- nominally 10 years of each experiment unless otherwise specified. We also show the sensitivities for 960 kT-MW-yr of DUNE, twice its nominal exposure, combined with 2.47 MT-MW-yr of HK.
}
\begin{tabular}{c|c|c|c}
$\delta=(-90^\circ,~0)$&no CID& 100\% eCID&100\% $\mu$CID\\
\hline
HK&$(20.8^\circ, 5.6^\circ)$&--&--\\
DUNE&$(17.8^\circ, 9.4^\circ)$&--&--\\
DUNE+HK&$(13.9^\circ,4.8^\circ)$&--&--\\
DUNE (20 yr)+HK &$(11.0^\circ, 4.5^\circ)$&--&--\\
DUNE+HK+NF(FNAL)&$(11.2^\circ,3.9^\circ)$&$(8.5^\circ, 3.2^\circ)$&$(9.0^\circ, 3.3^\circ)$\\
DUNE+HK+NF(BNL)&$(9.3^\circ, 3.9^\circ)$&$(8.0^\circ, 3.3^\circ)$&$(8.6^\circ, 3.4^\circ)$\\
\end{tabular}
\label{tab:results}
\end{table}

\begin{table}
\centering
\caption{The $1\sigma$ resolution on the other parameters for combinations of different experiments. For DUNE we assume 480 kT-MW-yr, for HK we assume 2.47 MT-MW-yr, for the neutrino factory exposure we assume  a 40 kT LAr detector and a total of $10^{22}$ muon decays -- nominally 10 years of each experiment unless otherwise specified. We also show the sensitivities for 960 kT-MW-yr of DUNE, twice its nominal exposure, combined with 2.47 MT-MW-yr of HK.
The current uncertainties are $\sim1^\circ$, $\sim0.1^\circ$, and $\sim20\times10^{-6}$ eV$^2$ for $\theta_{23}$, $\theta_{13}$, and $|\Delta m^2_{31}|$, respectively. We assume true NO.}
\begin{tabular}{c|c|c|c}
$\theta_{23}^\text{true}=(42^\circ,~45^\circ,~49^\circ)$&no CID& 100\% eCID&100\% $\mu$CID\\
\hline
HK&$(0.22^\circ, 1.37^\circ, 0.23^\circ)$&--&--\\
DUNE&$(0.33^\circ,1.32^\circ 0.38^\circ)$&--&--\\
DUNE+HK&$(0.18^\circ,1.04^\circ,0.20^\circ)$&--&--\\
DUNE (20 yr)+HK&$(0.15^\circ,0.92^\circ,0.16^\circ)$&--&--\\

DUNE+HK+NF(FNAL)&$(0.16^\circ,0.80^\circ,0.18^\circ)$&$(0.16^\circ,0.74^\circ, 0.18^\circ)$&$(0.15^\circ,0.70^\circ,0.17^\circ)$\\
DUNE+HK+NF(BNL)&$(0.16^\circ,0.56^\circ, 0.17^\circ)$&$(0.15^\circ,0.52^\circ, 0.17^\circ)$&$(0.14^\circ,0.52^\circ, 0.16^\circ)$\\
\hline\hline

$\theta_{13}^\text{true}=8.54^\circ$&& &\\
\hline
HK&$0.22^\circ$&--&--\\
DUNE&$0.21^\circ$&--&--\\
DUNE+HK&$0.16^\circ$&--&--\\
DUNE (20 yr)+HK&$0.15^\circ$&--&--\\
DUNE+HK+NF(FNAL)&$0.13^\circ$&$0.11^\circ$&$0.11^\circ$\\
DUNE+HK+NF(BNL)&$0.13^\circ$&$0.11^\circ$&$0.11^\circ$\\
\hline\hline

$(\Delta m_{31}^2)^\text{true}=2.511\cdot 10^{-3}~\text{eV}^2$&$[10^{-6}~\text{eV}^2]$&$[10^{-6}~\text{eV}^2]$ &$[10^{-6}~\text{eV}^2]$\\
\hline
HK&$9.8$&--&--\\
DUNE&$10.0$&--&--\\
DUNE+HK&$7.2$&--&--\\
DUNE (20 yr)+HK&$6.1$&--&--\\
DUNE+HK+NF(FNAL)&$6.0$&$5.8$&$6.0$\\
DUNE+HK+NF(BNL)&$5.6$&$5.5$&$5.3$\\
\end{tabular}
\label{tab:results_others}
\end{table}

\section{Conclusions}
Motivated by the recent P5 report, we study the physics potential of a neutrino factory in improving the precision on the determination of the remaining undetermined neutrino oscillation parameters with a focus on the complex phase $\delta$ which governs the amount of CP violation in the neutrino sector. We consider two different baselines and muon energies with a LAr detector based at SURF. 
Similar baselines and energies were considered as a low-energy neutrino factory \cite{Geer:2007kn,Bross:2007ts}  appropriate to measure CPV for large values of $\theta_{13}$ \cite{Tang:2009wp,Dighe:2011pa,Ballett:2012rz,Christensen:2013va}.
Such a neutrino factory setup could be achieved with a proton beam with 2-4 MW \cite{Holtkamp:2000xn,Ozaki:2001bb,NeutrinoFactory:2004odt,Geer:2006tb,Ankenbrandt:2009zza}.

We find that a neutrino factory has several important benefits for the community:
\begin{itemize}
\item Improved precision on several fundamental parameters including the amount of CP violation,
\item Improved flavor model differentiation capabilities,
\item A technological stepping stone on the way to a high energy muon collider to measure the electroweak sector with increased precision.
\end{itemize}
Such a machine may also be necessary in the event of a tension in the future oscillation data, a scenario that currently exists with, albeit at low significance.
As there are numerous valuable benefits to such a machine, we feel that it would be a strong addition to the neutrino oscillation experimental landscape moving forward.

We also investigated the role of charge identification of electrons or muons (or both).
We found that CID can lead to some improvement in the sensitivity to CP violation and the other parameters.
The impact is largest with larger statistics as CID helps to cut through systematics which become relevant at high statistics.
We also found that perfect CID is not required to get the full benefit as significant enhancement occurs at potentially realistic LArTPC CID as well.

A neutrino factory is a natural successor to the upcoming accelerator based long-baseline neutrino oscillation program to begin to move the neutrino sector towards the levels of precision already achieved in other sectors of particle physics.

\section*{Acknowledgements}
PBD acknowledges support by the United States Department of Energy under Grant Contract No.~DE-SC0012704. JG acknowledges support by the U.S. Department of Energy Office of Science under award number DE-SC0025448.

\appendix
\section{Number of events}
\label{sec:events}
\begin{figure}
\centering
\includegraphics[width=0.49\columnwidth]{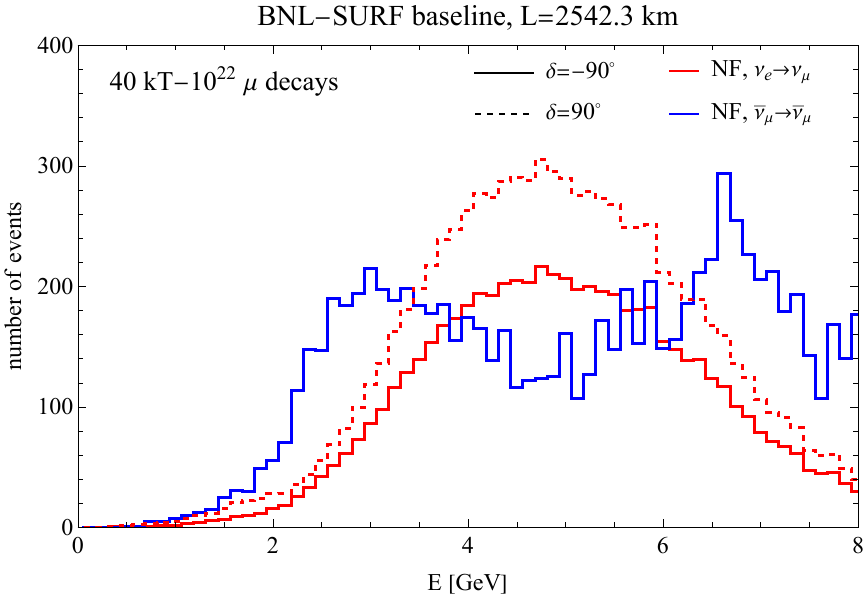}
\includegraphics[width=0.49\columnwidth]{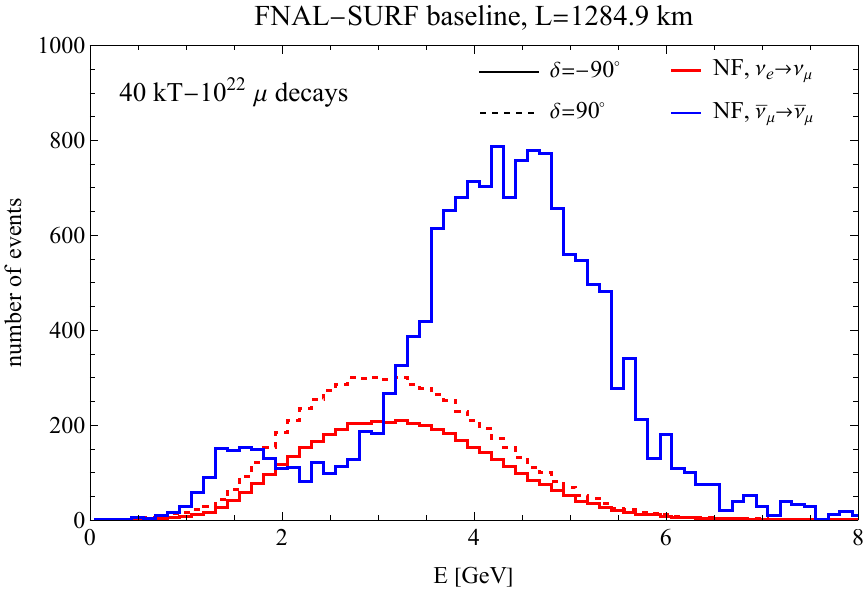}
\caption{Number of events at a neutrino factory for the BNL-SURF and FNAL-SURF baseline with muon energy 8 GeV and 5 GeV for $10^{22}$ muon decays (nominally 10 years). We show the number of events from the $
\nu_e\to \nu_\mu$ channel in red for varying values of $\delta$ and the dominant background from $\bar{\nu}_\mu \to\bar{\nu}_\mu$.}
\label{fig:numevents}
\end{figure}
In fig.~\ref{fig:numevents}
we show the number of events at a neutrino factory and DUNE. The background for the FNAL baseline is larger whereas the signal at the BNL baseline is where the background has a dip.
We also note that the shape of the disappearance spectra is quite different than that from a proton fixed target source due to the very different spectral shape, see fig.~\ref{fig:fluxes}.

\section{Results for inverted ordering and higher energies for BNL setup}
\label{sec:BNLres}
In tab.~\ref{tab:resultshigherE} we summarize the precision on the oscillation parameters for the BNL setup with higher muon energies of 12 GeV and 15 Gev. 

In tab.~\ref{tab:resultsIO} we show the precision for the oscillation parameters in inverted ordering.

\begin{figure*}
\centering
\includegraphics[width=0.49\columnwidth]{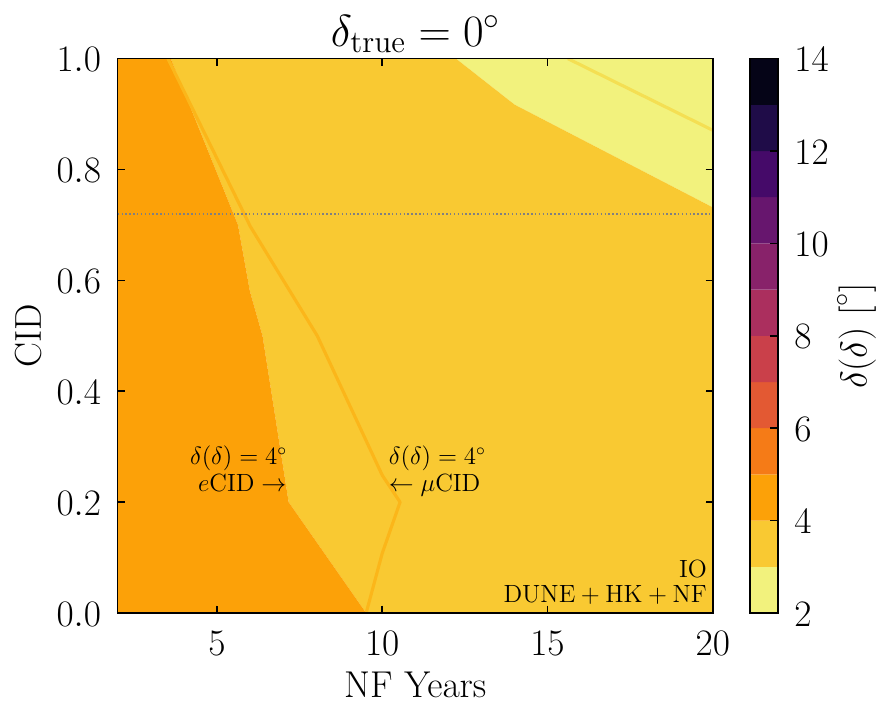}
\includegraphics[width=0.49\columnwidth]{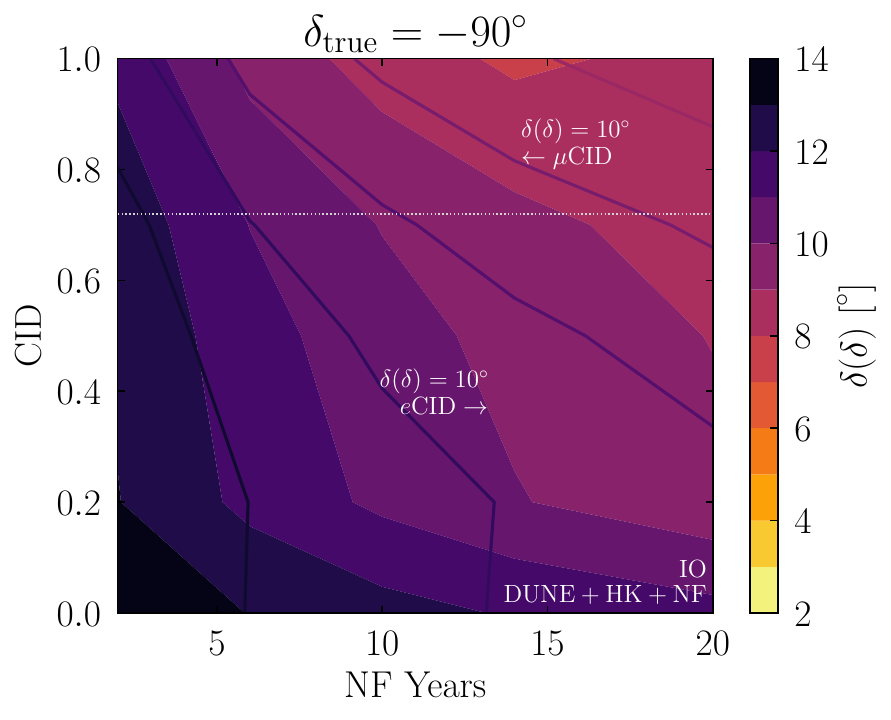}
\caption{Precision on $\delta$ in degrees as a function of neutrino factory exposure and CID in IO. We assume true IO.
The neutrino factory has a baseline and muon energy of 1284.9 km and 5 GeV and also includes 10 years of both DUNE and HK.
The shaded regions are for electron CID and the lines are for muon CID; electron CID does slightly better in most cases.
The horizontal lines show possibly achievable muon CID.
The true values of $\delta$ are $0$ (\textbf{left}) and $-90^\circ$ (\textbf{right}).
We assume as ratio of neutrino to anti-neutrino running time of 1:1 for the neutrino factory and DUNE 1:1 and 1:3 for HK.}
\label{fig:precision_deltaIO}
\end{figure*}

\begin{figure}
\centering
\includegraphics[width=0.45\columnwidth]{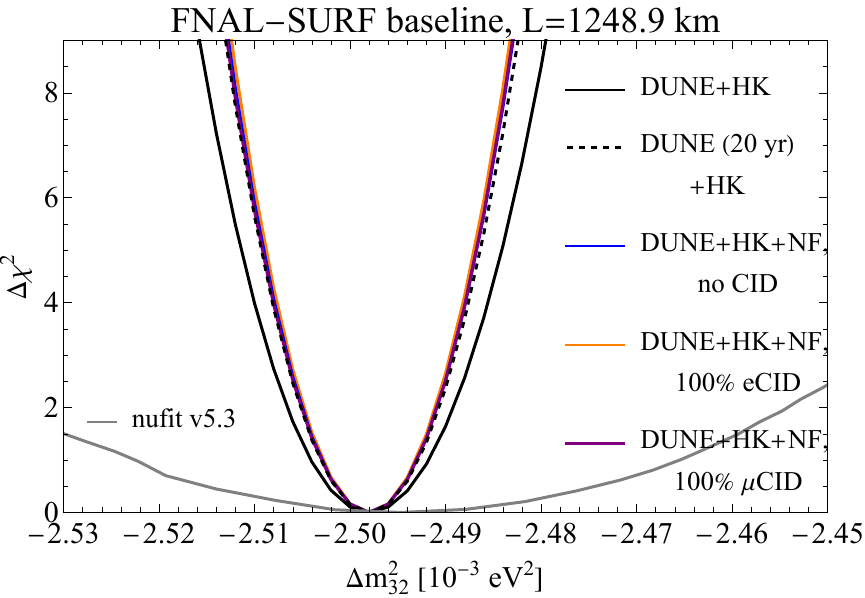}
\includegraphics[width=0.45\columnwidth]{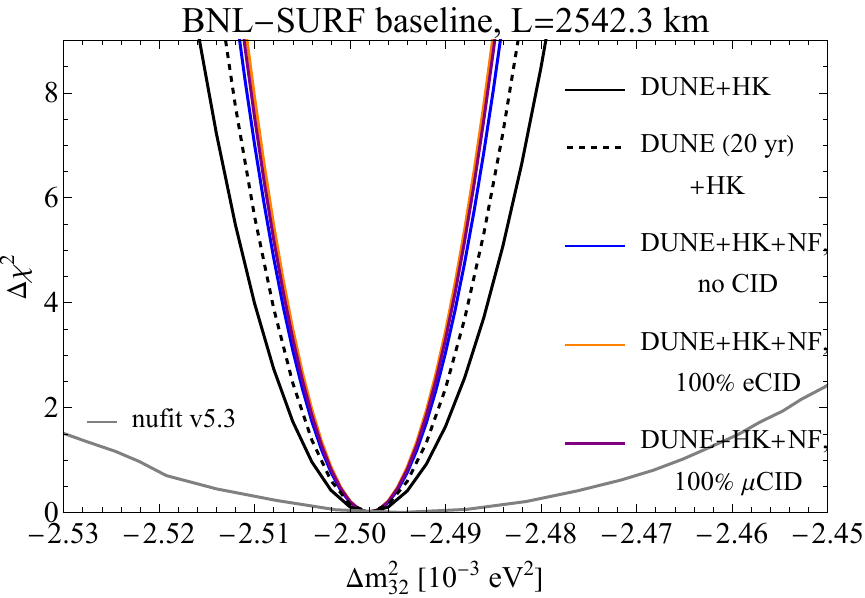}
\includegraphics[width=0.45\columnwidth]{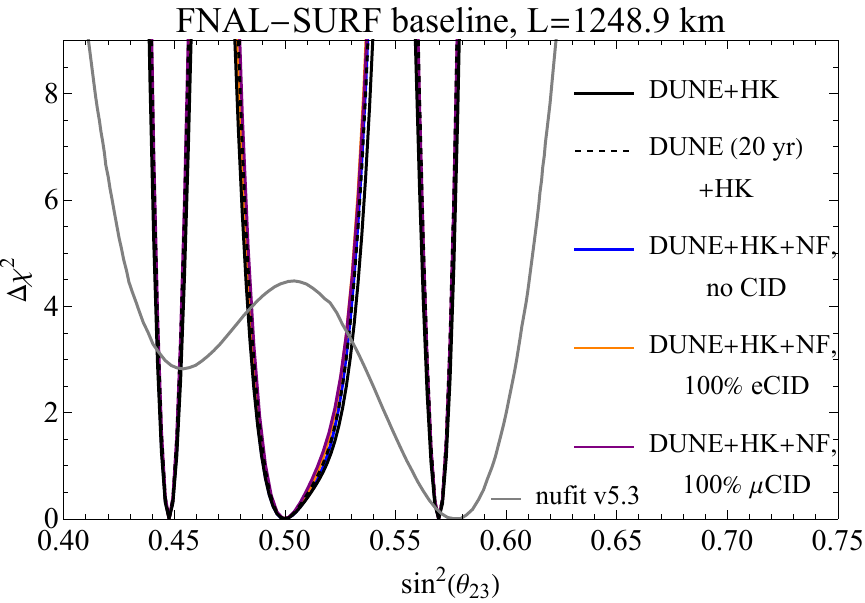}
\includegraphics[width=0.45\columnwidth]{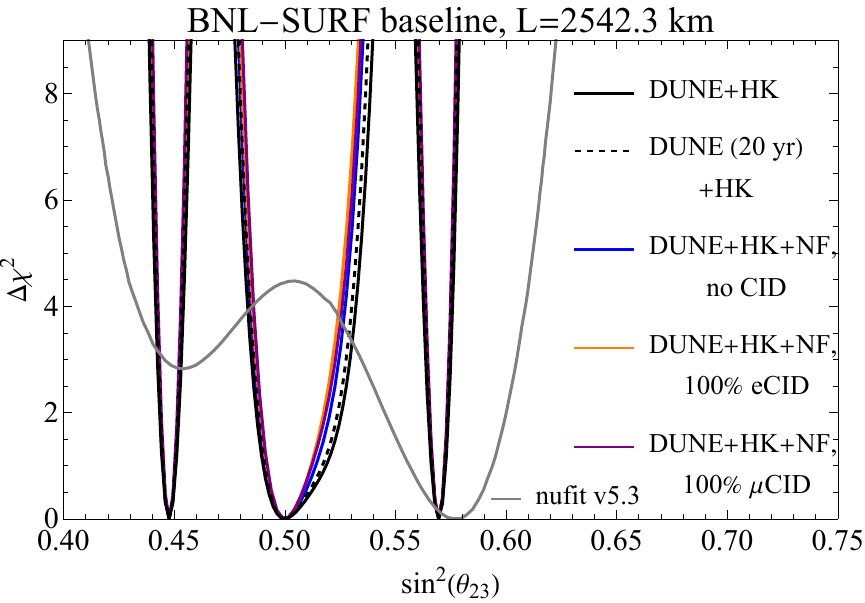}
\includegraphics[width=0.45\columnwidth]{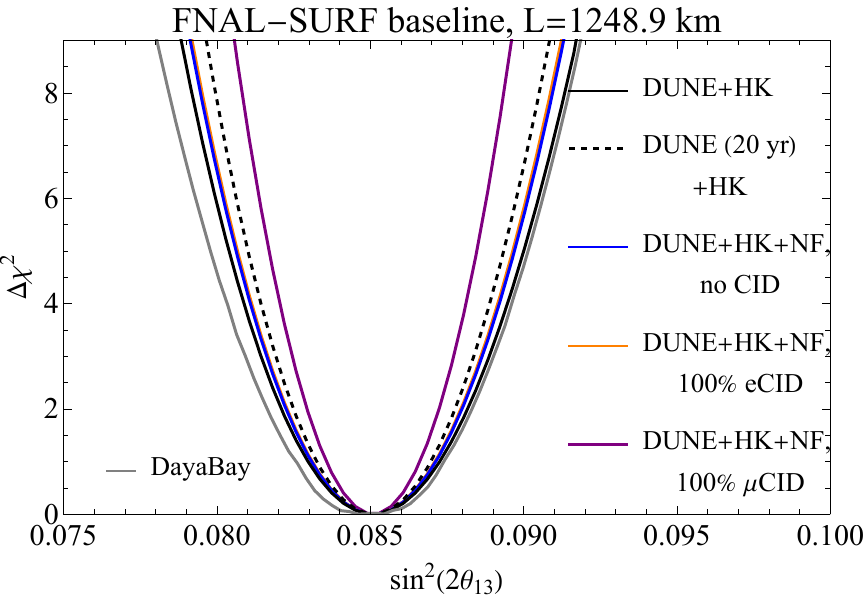}
\includegraphics[width=0.45\columnwidth]{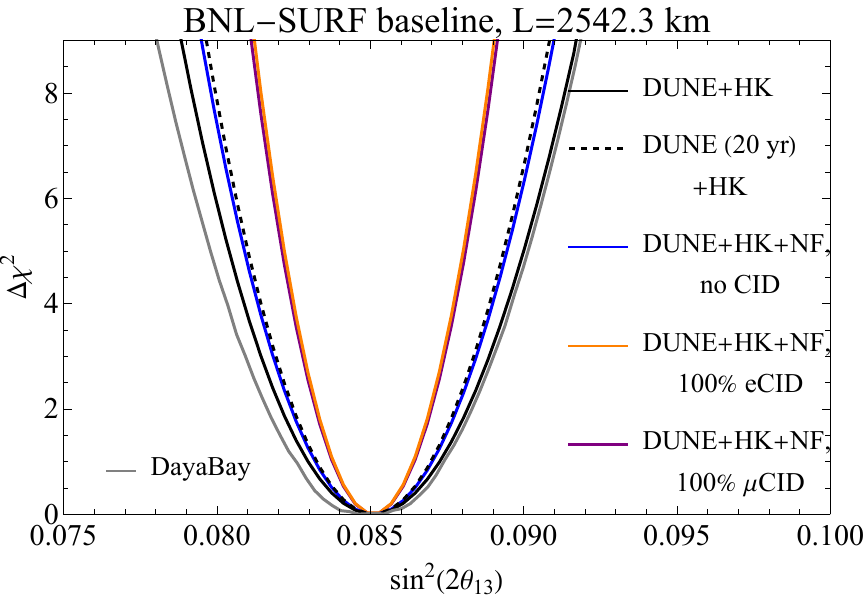}
\caption{Precision on $\Delta m_{31}^2, ~\sin^2(2\theta_{13}),~\sin^2 \theta_{23}$ for true IO. For $\theta_{23}$ we assume three different true values $\theta_{23}^\text{true}=(42^\circ,~45^\circ,~49^\circ)$. We show the results for 10 years of NF with $10^{21}$ muon decays per year and a 40 kT FD and $E_\mu=5 ~(8)$ GeV for the FNAL (BNL) setup on the left (right) plots, and for different assumptions on CID. The results for the NF also include 10 years of both HK and DUNE. 
We also compare the results to 20 years of DUNE in addition to 10 years of HK. The gray curves show the current precision on these parameters from \cite{Esteban:2020cvm,DayaBay:2022orm}.}
\label{fig:otherparametersIO}
\end{figure}

\begin{figure}
\centering
\includegraphics[width=0.47\columnwidth]{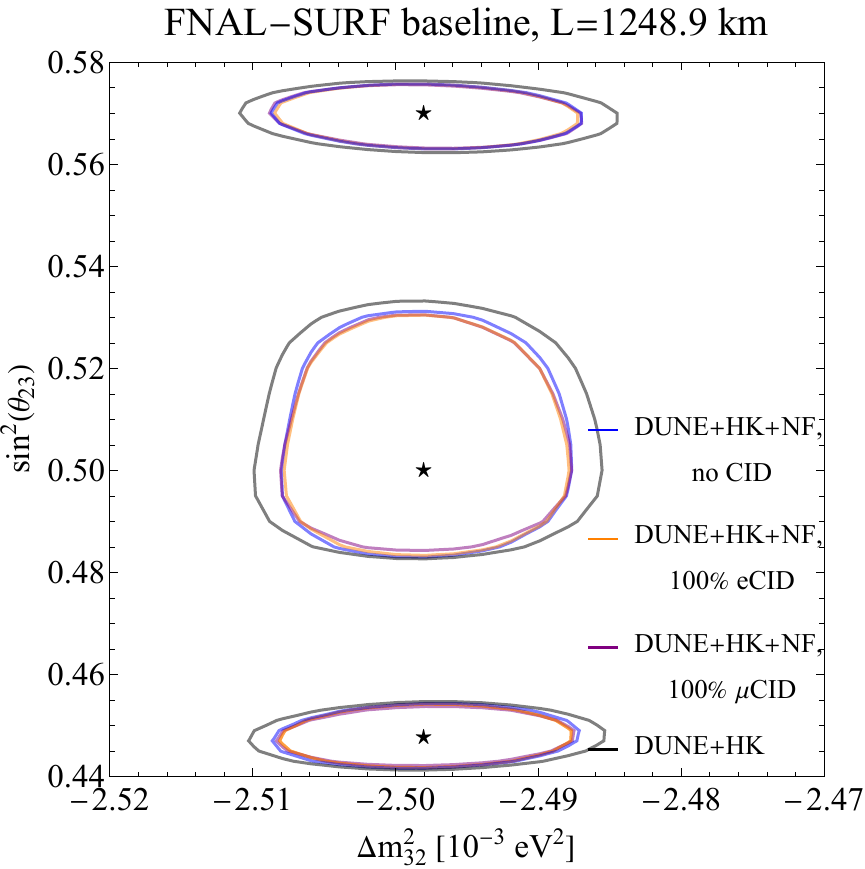}
\includegraphics[width=0.45\columnwidth]{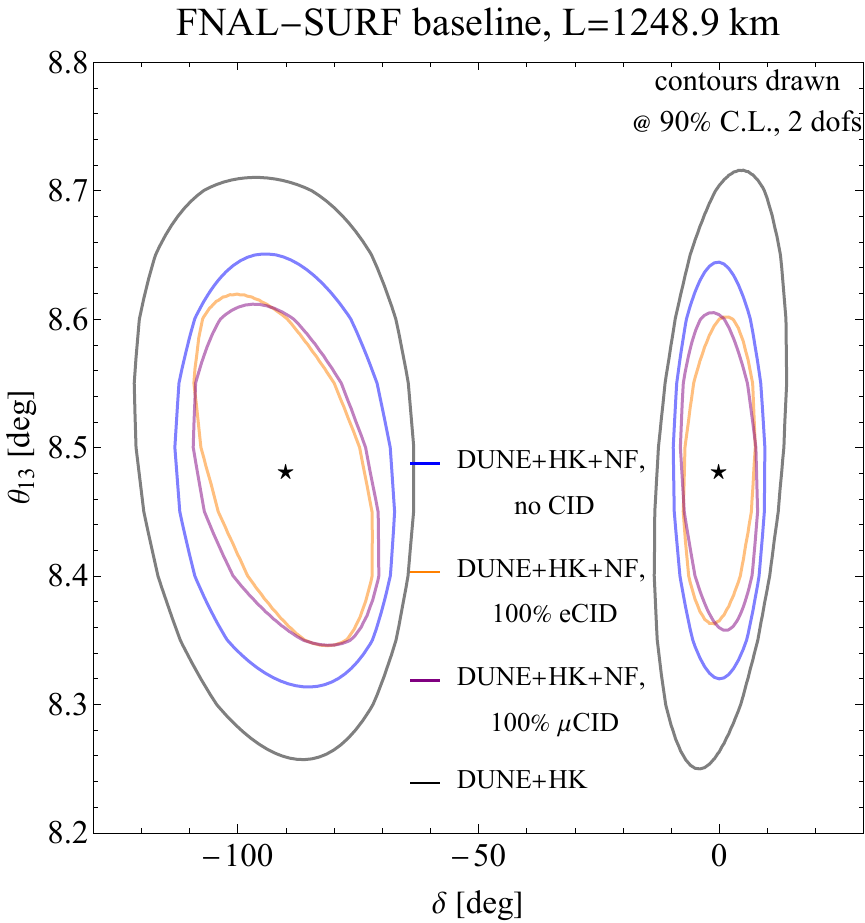}
\caption{Two-dimensional precision on the oscillation parameters assuming DUNE+HK+NF (10 years of each) with the FNAL-SURF baseline assuming different true values marked with a star. The contours are drawn at 90\% C.L. for 2 dofs. We assume true IO. }
\label{fig:otherparameters2dIO}
\end{figure}

In figs.~\ref{fig:precision_deltaIO}, \ref{fig:otherparametersIO}, \ref{fig:otherparameters2dIO} we show the corresponding results in IO to figs.~\ref{fig:precision_delta}, \ref{fig:otherparameters}, \ref{fig:otherparameters2d} in the main text.
The results are quite similar to the NO.

\begin{table}
\centering
\caption{Results for precision on parameters for BNL baseline at higher energies and true NO. For DUNE we assume 480 kT-MW-yr, for HK we assume 2.47 MT-MW-yr, for the neutrino factory exposure we assume a 40 kT LAr detector and a total of $10^{22}$ muon decays, nominally 10 years of each experiment.}
\begin{tabular}{c|c|c|c}
$\delta^\text{true}=(-90^\circ,~0)$&no CID& 100\% eCID&100\% $\mu$CID\\
\hline
DUNE+HK+NF(BNL) E = 12 GeV&$(10.8^\circ, 4.1^\circ)$&$(7.9^\circ, 3.4^\circ)$&$(8.6^\circ, 3.4^\circ)$\\
DUNE+HK+NF(BNL) E = 15 GeV&$(11.2^\circ, 4.2^\circ)$&$(8.0^\circ, 3.7^\circ)$&$(8.5^\circ, 3.6^\circ)$\\
\hline\hline
$\theta_{13}^\text{true}=8.54^\circ$&&&\\\hline
DUNE+HK+NF(BNL) E = 12 GeV&$0.13^\circ$&$0.12^\circ$&$0.12^\circ$\\
DUNE+HK+NF(BNL) E = 15 GeV&$0.13^\circ$&$0.12^\circ$&$0.12^\circ$\\

\hline\hline
$(\Delta m_{31}^2)^\text{true}=2.511\cdot 10^{-3}~\text{eV}^2$&$[10^{-6}\text{eV}^2]$&$[10^{-6}~\text{eV}^2]$&$[10^{-6}~\text{eV}^2]$
\\\hline
DUNE+HK+NF(BNL) E = 12  GeV&6.2&6.0&6.2\\
DUNE+HK+NF(BNL) E = 15 GeV&$6.4$&$6.1$&6.3\\
\hline\hline
$\theta_{23}^\text{true}=(42^\circ,~45^\circ, ~49^\circ)$&&&\\\hline
DUNE+HK+NF(BNL) E = 12  GeV&$(0.17^\circ,0.67^\circ,0.18^\circ)$&$(0.16^
\circ,0.61^\circ,0.18^\circ)$&$(0.15^\circ,0.61^\circ,0.18^\circ)$\\
DUNE+HK+NF(BNL) E = 15 GeV&$(0.17^\circ,0.73^\circ,0.18^\circ)$&$(0.16^\circ,0.65^\circ,0.18^\circ)$&$(0.16^\circ,0.67^\circ,0.18^\circ)$\\
\end{tabular}
\label{tab:resultshigherE}
\end{table}

\begin{table}
\centering
\caption{Resolution on the other parameters for combinations of different experiments assuming true inverted ordering. For DUNE we assume 480 kT-MW-yr, for HK we assume 2.47 MT-MW-yr, for the neutrino factory exposure we assume  a 40 kT LAr detector and a total of $10^{22}$ muon decays -- nominally 10 years of each experiment unless otherwise specified. We also show the sensitivities for 960 kT-MW-yr of DUNE, twice its nominal exposure, combined with 2.47 MT-MW-yr of HK.}
\begin{tabular}{c|c|c|c}
$\delta^\text{true}=(-90^\circ,~0)$&no CID& 100\% eCID&100\% $\mu$CID\\
\hline
HK&$(20.2^\circ,5.6^\circ)$&--&--\\
DUNE&$(18.8^\circ,8.6^\circ)$&--&--\\
DUNE+HK&$(14.2^\circ,4.7^\circ)$&--&--\\
DUNE (20 yr)+HK&$(11.4^\circ,4.3^\circ)$&--&--\\

DUNE+HK+NF(FNAL)&$(12.4^\circ,4.0^\circ)$&$(9.7^\circ,3.2^\circ)$&$(9.8^\circ,3.2^\circ)$\\
DUNE+HK+NF(BNL)&$(11.9^\circ,4.2^\circ)$&$(9.4^\circ,3.5^\circ)$&$(10.0^\circ,3.6^\circ)$\\

DUNE+HK+NF(BNL) $E=12$ GeV&$(12.0^\circ,4.3^\circ)$&$(8.8^\circ,3.7^\circ)$&$(9.2^\circ,3.7^\circ)$\\
DUNE+HK+NF(BNL) $E=15$ GeV&$(12.5^\circ,4.4^\circ)$&$(8.6^\circ,3.9^\circ)$&$(8.8^\circ,3.8^\circ)$\\
\hline\hline
$\theta_{13}^\text{true}=8.54^\circ$&&&\\\hline
HK&$0.22^\circ$&--&--\\
DUNE&$0.22^\circ$&--&--\\
DUNE+HK&$0.17^\circ$&--&--\\
DUNE (20 yr)+HK&$0.15^\circ$&--&--\\

DUNE+HK+NF(FNAL)&$0.16^\circ$&$0.14^\circ$&$0.13^\circ$\\
DUNE+HK+NF(BNL)&$0.16^\circ$&$0.12^\circ$&$0.12^\circ$\\

DUNE+HK+NF(BNL) $E=12$ GeV&$0.15^\circ$&$0.13^\circ$&$0.13^\circ$\\
DUNE+HK+NF(BNL) $E=15$ GeV&$0.16^\circ$&$0.14^\circ$&$0.14^\circ$\\

\hline\hline
$(\Delta m_{32}^2)^\text{true}=-2.498\cdot 10^{-3}~\text{eV}^2$&$[10^{-6}\text{eV}^2]$&$[10^{-6}~\text{eV}^2]$&$[10^{-6}~\text{eV}^2]$
\\\hline
HK&$8.9$&--&--\\
DUNE&$8.9$&--&--\\
DUNE+HK&$6.2$&--&--\\
DUNE (20 yr)+HK&$5.1$&--&--\\

DUNE+HK+NF(FNAL)&$5.0$&$4.8$&$5.0$\\
DUNE+HK+NF(BNL)&$4.5$&$4.3$&$4.4$\\

DUNE+HK+NF(BNL) $E=12$ GeV&$5.1$&$5.0$&$5.0$\\
DUNE+HK+NF(BNL) $E=15$ GeV&$5.4$&$5.0$&$5.1$\\
\hline\hline
$\theta_{23}^\text{true}=(42^\circ,~45^\circ, ~49^\circ)$&&&\\\hline
HK&$(0.22^\circ,1.33^\circ,0.23^\circ)$&--&--\\
DUNE&$(0.35^\circ,1.32^\circ,0.35^\circ)$&--&--\\
DUNE+HK&$(0.18^\circ,1.07^\circ,0.19^\circ)$&--&--\\
DUNE (20 yr)+HK&$(0.16^\circ,0.97^\circ,0.17^\circ)$&--&--\\

DUNE+HK+NF(FNAL)&$(0.17^\circ,1.01^\circ,0.17^\circ)$&$(0.17^\circ,0.91^\circ,0.17^\circ)$&$(0.16^\circ,0.86^\circ,0.17^\circ)$\\
DUNE+HK+NF(BNL)&$(0.17^\circ,0.81^\circ,0.17^\circ)$&$(0.16^\circ,0.68^\circ,0.17^\circ)$&$(0.15^\circ,0.70^\circ,0.16^\circ)$\\

DUNE+HK+NF(BNL) $E=12$ GeV&$(0.17^\circ,0.97^\circ,0.17^\circ)$&$(0.17^\circ,0.78^\circ,0.17^\circ)$&$(0.17^\circ,0.86^\circ,0.18^\circ)$\\
DUNE+HK+NF(BNL) $E=15$ GeV&$(0.18^\circ,1.00^\circ,0.18^\circ)$&$(0.17^\circ,0.84^\circ,0.18^\circ)$&$(0.17^\circ,0.91^\circ,0.18^\circ)$\\
\end{tabular}
\label{tab:resultsIO}
\end{table}

\bibliographystyle{JHEP}
\bibliography{main}

\end{document}